\documentclass[conference]{IEEEtran}
\IEEEoverridecommandlockouts
\usepackage{cite}
\usepackage{amsmath,amssymb,amsfonts}
\usepackage{algorithmic}
\usepackage{graphicx}
\usepackage{textcomp}
\usepackage{hyperref}
\usepackage[table]{xcolor}
\usepackage[pass]{geometry}
\usepackage{booktabs}
\usepackage{amsmath}
\usepackage{mathtools}
\usepackage{diagbox}
\usepackage{tabularx}
\usepackage{threeparttable}
\usepackage{siunitx}
\usepackage{eurosym}
\usepackage{multirow}
\usepackage{caption}

\definecolor{gr}{rgb}{0.67, 0.88, 0.69}
\definecolor{or}{rgb}{1.0, 0.65, 0.0}
\definecolor{re}{rgb}{1.0, 0.41, 0.38}

\definecolor{darkcerulean}{rgb}{0.03, 0.27, 0.49}
\definecolor{ao}{rgb}{0.0, 0.5, 0.0}
\definecolor{bred}{rgb}{0.8, 0.25, 0.33}
\definecolor{crimson}{rgb}{0.86, 0.08, 0.24}

\newcolumntype{Y}{>{\centering\arraybackslash}X}

\def\BibTeX{{\rm B\kern-.05em{\sc i\kern-.025em b}\kern-.08em
    T\kern-.1667em\lower.7ex\hbox{E}\kern-.125emX}}

\begin{document}

\title{Model Reduction in Capacity Expansion Planning Problems via Renewable Generation Site Selection}

\author{\IEEEauthorblockN{David Radu, Antoine Dubois, Mathias Berger, Damien Ernst}
\IEEEauthorblockA{\textit{Dept. of Electrical Engineering and Computer Science, University of Liège, 4000 Liège, Belgium} \\
\{dcradu, antoine.dubois, mathias.berger, dernst\}@uliege.be}}

\IEEEoverridecommandlockouts
\IEEEpubid{\makebox[\columnwidth]{978-1-6654-3597-0/21/\$31.00~\copyright2021 IEEE \hfill} \hspace{\columnsep}\makebox[\columnwidth]{ }}

\maketitle

\IEEEpubidadjcol

\begin{abstract}
The accurate representation of variable renewable generation (RES, e.g., wind, solar PV) assets in capacity expansion planning (CEP) studies is paramount to capture spatial and temporal correlations that may exist between sites and impact both power system design and operation. However, it typically has a high computational cost. This paper proposes a method to reduce the spatial dimension of CEP problems while preserving an accurate representation of renewable energy sources. A two-stage approach is proposed to this end. In the first stage, relevant sites are identified via a screening routine that discards the locations with little impact on system design. In the second stage, the subset of relevant RES sites previously identified is used in a CEP problem to determine the optimal configuration of the power system. The proposed method is tested on a realistic EU case study and its performance is benchmarked against a CEP set-up in which the entire set of candidate RES sites is available. The method shows great promise, with the screening stage consistently identifying 90\% of the optimal RES sites while discarding up to 54\% of the total number of candidate locations. This leads to a peak memory reduction of up to 41\% and solver runtime gains between 31\% and 46\%, depending on the weather year considered.
\end{abstract}


\section{Introduction}


Capacity expansion planning (CEP) problems are powerful tools for the design, analysis and implementation of energy system decarbonisation policies. In such frameworks, the accurate spatiotemporal representation of variable renewable energy generation (RES, e.g., wind, solar PV) is paramount for the precise estimation of capacity requirements \cite{Pfenninger2014}. However, the detailed modelling of RES comes at a high computational cost and ways to mitigate this issue in order to strike the right balance between accuracy and computational effort when solving such problems are necessary, yet seldom proposed. For example, a highly detailed representation of RES within a CEP set-up cast as a linear program (LP) is proposed by MacDonald et al. \cite{MacDonald2016}, yet the reported runtimes (thousands of core hours for large-scale instances) limit its use in practice and its reproducibility. Wu et al. \cite{Callaway2017} also propose an LP-cast CEP framework in which high-resolution RES modelling is made possible via a GIS-based resource assessment tool. Nonetheless, the coefficient matrix stores hourly capacity factor values at each location and is therefore full, which limits the scalability of the proposed method to a few hundreds of candidate RES sites only, thus rendering it unsuitable for large-scale applications. 

Although plenty of work has been carried out in recent years to develop temporal reduction techniques for RES in CEP settings \cite{Hoffmann2020}, studies tackling the issue of spatial model reduction are scarce. Cohen et al. \cite{Cohen2018} suggest the aggregation of RES in resource regions, with wind and solar PV resources over the contiguous United States being modelled via 356 and 134 profiles, respectively. In a similar vein, Hörsch and Brown \cite{Horsch2017} leverage a CEP framework formulated as an LP to assess the impact of spatial resolution on the outcomes of co-optimizing generation and transmission assets across Europe. A network reduction process based on k-means clustering is incorporated in their method and the resulting topology serves as the basis for modelling renewable resources. More precisely, Europe-wide RES are represented via 37 to 362 different aggregate profiles, depending on the desired number of network clusters. While spatial aggregation approaches, as the ones proposed in \cite{Cohen2018, Horsch2017}, partly mitigate the aforementioned computational issues \cite{MacDonald2016, Callaway2017}, the limited number of RES profiles considered hinders their ability to exploit the benefits of resource diversity which, in turn, can lead to system cost overestimation \cite{Frew2016}.

This paper proposes a method to reduce the spatial dimension and decrease the computational requirements of CEP problems while preserving a detailed representation of RES assets. This is achieved by leveraging a two-stage heuristic that can be described as follows. The first stage, which is cast as an LP, is used to screen a set of candidate sites and identify sites that have little impact on optimal system design, which are then discarded. In the second stage, information (geo-positioning and capacity factors time series) about the remaining sites is used as input data in a CEP framework that determines the installed capacities of generation, storage and transmission assets leading to a minimum-cost system configuration. Thus, the proposed method makes it possible to reduce the size of the CEP problem, and therefore enables memory and computation time savings.

The paper is structured as follows. Section \ref{methods} details the methods at the core of the proposed two-stage approach. Then, Section \ref{casestudy} briefly describes the case study used to showcase the applicability of the suggested approach before results are reported in Section \ref{results}. Section \ref{conclusion} concludes the paper and discusses future work avenues.

\section{Method}\label{methods}

The proposed solution method (or \texttt{SM}) is introduced in this section. Firstly, the standard CEP framework (from hereon, the \texttt{FLP}) is formulated. In the remainder of this paper, the \texttt{FLP} denotes the CEP set-up that simultaneously tackles the siting and sizing of RES assets, as well as the sizing of other power system (e.g., generation, storage or transmission) technologies. Then, the screening method for candidate RES sites (\texttt{SITE}) that enables the formulation of a reduced-size CEP framework (from hereon, the \texttt{RLP}) is described. The \texttt{SITE}-\texttt{RLP} sequence will hereafter be referred to as the \texttt{SM}.

\subsection{Capacity expansion planning framework}

Let $\mathcal{N}_B$ and $\mathcal{L}$ be the sets of existing buses and transmission corridors, respectively. Let $\mathcal{N}_R$ be a set of candidate RES sites that may be connected to buses $n \in \mathcal{N}_B$, which is partitioned into disjoint subsets $\mathcal{N}_R^n$. The CEP formulation reads \vspace{-10pt}

\begin{align}
\underset{\tiny \begin{array}{cc} \mathbf{K}, (\mathbf{p}_t)_{t \in \mathcal{T}} \end{array}} \min \hspace{-2pt} &\omega \Big[\sum_{\substack{n \in \mathcal{N}_B \\ m \in \mathcal{N}_R^n}} \big(\zeta^{m} + \theta^m_f\big) K_{nm} + \sum_{\substack{n \in \mathcal{N}_B \\ j \in \mathcal{G}\cup\mathcal{S}}} \big(\zeta^{j} + \theta^j_f\big) K_{nj} \nonumber\\&+ \sum_{l \in \mathcal{L}} \big(\zeta^l + \theta^l_f\big) K_l\Big] + \sum_{t \in \mathcal{T}}\Big[ \sum_{l \in \mathcal{L}} \theta^l_v |p_{lt}| \tag{1a}\label{eq:objective}\\&+ \sum_{\substack{n \in \mathcal{N}_B \\ m \in \mathcal{N}_R^n}} \theta^m_v p_{nmt} + \sum_{\substack{n \in \mathcal{N}_B \\ j \in \mathcal{G}\cup\mathcal{S}}} \theta^j_v |p_{njt}| + \sum_{n \in \mathcal{N}_B} \theta^{e} p^{e}_{nt}\Big] \nonumber
\end{align}
\vspace{-10pt}
\allowdisplaybreaks
\begin{subequations}
\begin{align}
\text{s.t.} &\sum_{\substack{m \in \mathcal{N}_R^n}} p_{nmt} + \sum_{g \in \mathcal{G}} p_{ngt} + \sum_{s \in \mathcal{S}} p^D_{nst} + \sum_{l \in \mathcal{L}_n^{+}} p_{lt} + p^{e}_{nt} \nonumber\\ & \hspace{1mm} = \lambda_{nt} + \sum_{s \in \mathcal{S}} p^C_{nst} + \sum_{l \in \mathcal{L}_n^{-}} p_{lt} , \mbox{ } \forall n \in \mathcal{N}_B, \forall t \in \mathcal{T}  \tag{1b}\label{eq:subeq_balance}\\[1pt]
&p_{nmt} \le \pi_{nmt} (\kappa^0_{nm} + K_{nm}), \nonumber\\ &\hspace{20mm} \forall n \in \mathcal{N}_B, \forall m \in \mathcal{N}_R^n, \forall t \in \mathcal{T} \tag{1c}\label{eq:subeq_res_flow}\\[1pt]
&\kappa^0_{nm} + K_{nm} \le \Bar{\kappa}_{nm}, \mbox{ } \forall n \in \mathcal{N}_B, \forall m \in \mathcal{N}_R^n \tag{1d}\label{eq:subeq_res_cap} \\[1pt]
&p_{ngt} \le \kappa^0_{ng} + K_{ng}, \mbox{ } \forall n \in \mathcal{N}_B, \forall g \in \mathcal{G}, \forall t \in \mathcal{T} \tag{1e}\label{eq:subeq_disp_flow} \\[1pt]
&\kappa^0_{ng} + K_{ng} \le \Bar{\kappa}_{ng}, \mbox{ } \forall n \in \mathcal{N}_B, \forall g \in \mathcal{G} \tag{1f}\label{eq:subeq_disp_cap} \\[1pt]
&p_{nst} = -p^C_{nst}+p^D_{nst}, \forall n \in \mathcal{N}_B, \forall s \in \mathcal{S}, \forall t \in \mathcal{T} \tag{1g}\label{eq:subeq_sto_def}\\[1pt]
&|p_{nst}| \le \phi_s (\kappa^0_{ns} + K_{ns}), \forall n \in \mathcal{N}_B, \forall s \in \mathcal{S}, \forall t \in \mathcal{T} \tag{1h}\label{eq:subeq_sto_ch}\\[1pt]
&e_{nst} \le \kappa^0_{ns} + K_{ns}, \tag{1i}\label{eq:subeq_sto_soc_lim} \forall n \in \mathcal{N}_B, \forall s \in \mathcal{S}, \forall t \in \mathcal{T}\\[3pt]
&e_{nst} = \eta^{SD}_s e_{ns(t-1)} + \eta^{C}_s p^C_{nst} - p^D_{nst}/\eta^D_{s}, \nonumber\\ & \hspace{35mm} \forall n \in \mathcal{N}_B, \forall s \in \mathcal{S}, \forall t \in \mathcal{T} \tag{1j}\label{eq:subeq_sto_soc}\\[1pt]
&\kappa^0_{ns} + K_{ns} \le \Bar{\kappa}_{ns}, \mbox{ } \forall n \in \mathcal{N}_B, \forall s \in \mathcal{S} \tag{1k}\label{eq:subeq_sto_cap} \\[1pt]
&|p_{lt}| \le \kappa^{0}_l + K_l, \mbox{ } \forall l \in \mathcal{L}, \forall t \in \mathcal{T} \tag{1l}\label{eq:subeq_tr_flow} \\[1pt]
&\kappa^{0}_l + K_l \le \Bar{\kappa}_l, \mbox{ } \forall l \in \mathcal{L} \tag{1m}\label{eq:subeq_tr_cap}
\end{align}
\end{subequations}

The problem described in (1a-m) minimizes total system cost subject to a set of constraints of the underlying assets. The objective function (\ref{eq:objective}) comprises capital expenditure, fixed and variable operating costs of the generation, storage and transmission assets, as well as the economic penalties associated with unserved demand. Constraint (1b) enforces the energy balance at each bus, while the operation and sizing of RES assets is modelled via (1c-d). Note that a single RES technology $r \in \mathcal{R}$ is associated with each site $m \in \mathcal{N}_R$. Then, conventional generators are modelled via (1e-f) and the operation and sizing of storage units follows (1g-k). Finally, constraints (1l-m) encode the transportation model governing the power flows in transmission links. It is worth noting that, although the absolute values in Eqs. (1a), (1h) or (1l) render the CEP problem described in (1a-m) non-linear, it can be cast as an LP using standard reformulation techniques.

\subsection{Renewable sites selection method} 

The proposed \texttt{SM} works by decoupling the siting and sizing of RES assets. At first, the \texttt{SITE} stage is leveraged to screen the sets of candidate RES locations and identify those sites that play a role in the optimal system design, while discarding the rest. To this end, the siting problem is formulated by i) discarding some complicating variables and approximating a subset of complicating constraints (i.e., the ones associated with dispatchable power generation, storage systems and power flows in transmission lines) and ii) relaxing and taking linear combinations, as well as scaling the right-hand site coefficients of certain equality constraints (i.e., the power balance equations). The objective function (\ref{eq:siting_obj}) is obtained by preserving the terms related to the costs of deploying and operating RES technologies and the economic penalty associated with unserved demand. Then, the constraints discarded from (1a-m) are approximated via two parameters found in (2b). More formally, let $\mathcal{T}$ be the set of time periods, let $\mathcal{T}_{\tau} \subseteq \mathcal{T}, \mbox{ } |\mathcal{T}_{\tau}|=\delta\tau, \mbox{ } \tau = 1, \ldots, T,$ be a collection of disjoint subsets forming a partition of $\mathcal{T}$ into time slices of length $\delta\tau$. More precisely, $\delta\tau$ represents the length of a time slice (e.g., one hour, one day) over which the energy balance in (2b) is enforced and its role is to emulate the behavior of storage assets shifting RES supply in time. Furthermore, let $\xi_{\tau}^n \in \mathbb{R}_+$ denote regional minimum RES feed-in targets enforced over every time slice $\mathcal{T}_\tau, \mbox{ } \tau = 1, \ldots, T$. This parameter enforces a minimum level of local power production from renewable sources which i) mirrors the effect of transmission constraints and ii) accounts for low-carbon legacy generation capacity that would offset the country-specific RES requirements. Constraints (1c-d) are preserved as such and the siting problem thus reads \vspace{-10pt}

\begin{align}
\underset{\tiny \begin{array}{cc} \mathbf{K}, (\mathbf{p}_t)_{t \in \mathcal{T}} \end{array}} \min \hspace{10pt} & \omega \Big[\sum_{\substack{n \in \mathcal{N}_B \\ m \in \mathcal{N}_R^n}} \big(\zeta^{m} + \theta^m_f\big) K_{nm}\Big] \mbox{ } + \nonumber\\ & \sum_{t \in \mathcal{T}} \Big[ \sum_{\substack{n \in \mathcal{N}_B \\ m \in \mathcal{N}_R^n}} \theta_v^r p_{nmt} + \sum_{n \in \mathcal{N}_B} \theta^{e} p^{e}_{nt} \Big] \label{eq:siting_obj} \tag{2a} \nonumber
\end{align}
\vspace{-10pt}
\begin{subequations}
\begin{align}
\text{s.t.} & \sum_{t \in \mathcal{T_{\tau}}} \Big[ \sum_{\substack{m \in \mathcal{N}_R^n}} p_{nmt} + p^{e}_{nt} \Big] \ge \xi_{\tau}^{n} \sum_{t \in \mathcal{T_{\tau}}} \lambda_{nt}, \nonumber\\ & \hspace{30mm} \forall n \in \mathcal{N}_B, \mbox{ } \forall \tau \in \{1, \ldots, T\} \tag{2b}\label{eq:en_balance_n}\\[1pt]
&p_{nmt} \le \pi_{nmt} (\kappa^0_{nm} + K_{nm}), \nonumber\\ & \hspace{30mm} \forall n \in \mathcal{N}_B, \forall m \in \mathcal{N}_R^n, \forall t \in \mathcal{T}\tag{2c}\label{eq:infeed_def}\\[1pt]
&\kappa^0_{nm} + K_{nm} \le \Bar{\kappa}_{nm}, \mbox{ } \forall n \in \mathcal{N}_B, \forall m \in \mathcal{N}_R^m \tag{2d}\label{eq:siting_cap}
\end{align}
\end{subequations}

For every $n \in \mathcal{N}_B$, the problem returns the set of candidate RES sites identified as relevant (with an installed capacity above \SI{1}{MW}) in the optimal system design, i.e. $\mathcal{N}_{\texttt{SITE}}^n$. Then, the \texttt{RLP} is built by replacing $\mathcal{N}_R^n$ with $\mathcal{N}_{\texttt{SITE}}^n$ in constraints (1a-d) of the CEP problem.

\section{Case Study}\label{casestudy}


\subsubsection*{Input Data}

The analysis is conducted for three individual weather years (i.e., 2016, '17 and '18) and over 33 countries within the \textit{ENTSO-E} system. The siting stage relies on hourly-sampled resource data obtained from the \textit{ERA5} reanalysis database \cite{ERA5} at a spatial resolution of \ang{1.0}. The mapping of resource data to capacity factors time series is achieved via the transfer functions of appropriate conversion equipment for each individual technology. More precisely, a site-specific selection of wind generators is carried out based on the \textit{IEC 61400} standard \cite{IEC61400} and four different converters are available for deployment (i.e., the \textit{Vestas V110, V90, V117} and \textit{V164}), each of them suitable for specific wind regimes. The selection of solar energy converters is done on a technology basis, with the \textit{TrinaSolar DEG15MC} module available for utility-scale PV deployment and the \textit{TrinaSolar DD06M} array available for distributed PV generation. A greenfield approach is adopted, i.e., no legacy capacity of RES assets is considered, while the technical potential is estimated via a land eligibility assessment framework \cite{Ryberg2017} that yields eligible surface areas for RES deployment for a set of 1740 candidate sites. A set of assumptions pertaining to the power densities of different generation technologies are then made to map surface areas into maximum allowable installed capacities, i.e., technical potentials. Specifically, a density of \SI{5}{MW/km^2} is considered for wind deployments \cite{WindEurope2020}. With respect to solar PV units, power densities of \SI{40}{MW/km^2} and \SI{16}{MW/km^2} are considered for utility-scale and residential installations, respectively \cite{Trondle2019}. Electricity demand time series for all considered countries are retrieved from the \textit{OPSD} platform \cite{OPSD}.

\begin{figure}
\centering
\includegraphics[width=.9\linewidth]{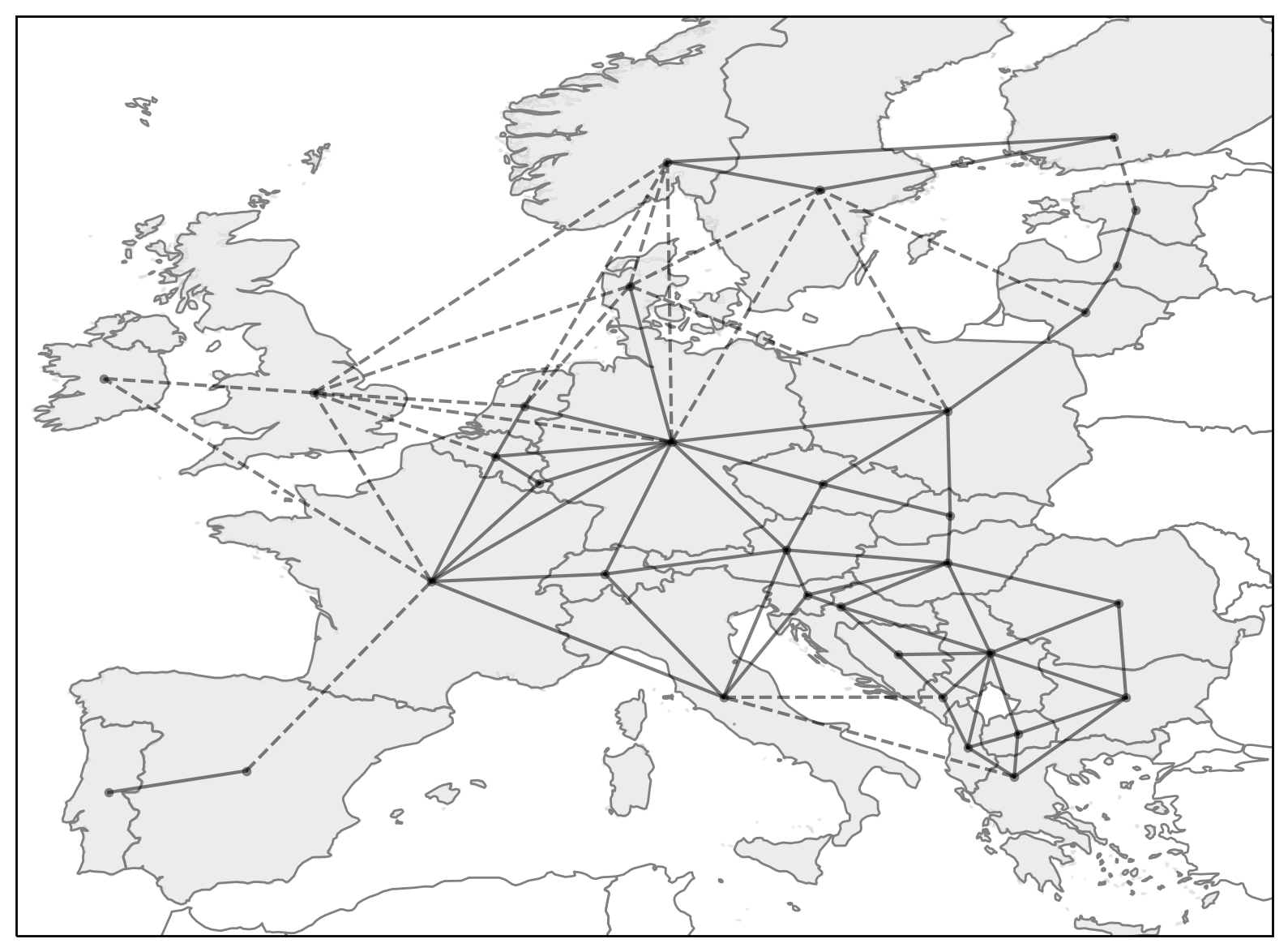}
\caption{System topology in the capacity expansion planning framework. AC connections displayed in full lines, DC links shown in dotted lines. The 33 nodes form the set $\mathcal{N}_B$ in (1a-m) and (2a-d).}
\label{fig:topology}
\end{figure}

The CEP frameworks (i.e., both the \texttt{FLP} and \texttt{RLP}) follow a centralized planning approach and build upon the 2018 TYNDP dataset, where each European country is modelled as one node \cite{TYNDP2018}. The resulting network topology is displayed in Fig. \ref{fig:topology}. In this exercise, the expansion of the transmission network is limited to the reinforcement of existing links. Furthermore, the total capacity of each link may not exceed twice the 2040 capacity estimated for this link in the TYNDP. Besides the four RES technologies sited in the previous stage, three more generation technologies are available for power generation, namely run-of-river (ROR) and reservoir-based (STO) hydro, as well as combined-cycle gas turbines (CCGT), with the latter being the only of the three that is also sized in (1a-m). The existing capacities of the other two are retrieved from \cite{JRC_HY}, where the existence of \SI{34}{GW} of ROR and \SI{98}{GW} of STO installations is reported. Then, two technologies are available for electricity storage, namely pumped-hydro (PHS) units and Li-Ion batteries. The latter is the only one being sized in (1a-m) and a fixed energy-to-power ratio of \SI{4}{h} is assumed. The legacy capacity of the former is retrieved from \cite{JRC_HY}, where \SI{55}{GW}/\SI{1950}{GWh} of PHS storage is reported. The CEP problem is implemented in PyPSA 0.17 \cite{PyPSA}, while the techno-economic assumptions are gathered in \cite{dox_repo_method}.

\subsubsection*{Parametrization of the \texttt{SITE} stage}

The two parameters of (2a-d) are defined as follows. First, the slicing period $\delta\tau$ is considered to be equal to \SI{24}{h}, which corresponds to the nonzero frequency component of the aggregate EU-wide RES capacity factor time series with the largest amplitude (i.e., as provided by a discrete Fourier transform). Then, the country-dependent $\xi_{\tau}^n$ values are assumed not to be time-dependent and their estimation proceeds as follows. First, the residual demand (i.e., the difference between demand and generation potential of legacy dispachable units) is computed at peak load conditions. Then, the RES generation potential during the same time instants is determined. For each country, if RES potential exceeds the electricity demand for at least half the time steps in the optimization horizon, its potential transmission capabilities (i.e., 2040 TYNDP capacity limits times the length of slicing period $\delta\tau$) are added to the residual demand, as the country is a potential exporter of electricity in the EU-wide system. Conversely, if the electricity demand is higher than the RES potential most of the time, the transmission capabilities of that country are subtracted from the residual demand, as cross-border exchanges will oftentimes be used to cover for the domestic electricity needs. Finally, the $\xi_{\tau}^n$ values are determined as the ratio between the RES potential and the transmission capacity-adjusted residual demand, respectively.

\subsubsection*{Implementation}

The \texttt{SM}, as well as the \texttt{FLP} are implemented in Python 3.7 and the proposed instances are run on a workstation running under CentOS, with an 18-core Intel Xeon Gold 6140 CPU clocking at \SI{2.3}{GHz} and \SI{256}{GB} RAM. Gurobi 9.0 was used to solve both (1a-m) and (2a-d). The dataset and code used in these simulations are available at \cite{dox_repo_method} and \cite{replan_git}.

\section{Results}\label{results}

The results of a set of experiments evaluating the performance of the \texttt{SM} against the \texttt{FLP} are detailed in this section.

Table \ref{tab:count} summarizes the performance of the siting stage by means of two indicators. First, the technology-specific spatial reduction share ($\gamma_r$) denotes the proportion of initial candidate RES sites discarded via \texttt{SITE}. Then, the screening accuracy ($\alpha_r$) measures the ability of the method to identify the relevant candidate RES sites. More formally, let $\mathcal{R}$ be the set of renewable technologies and let $\mathcal{N}_R^r$ be the subset of sites with technology $r \in \mathcal{R}$ (these subsets are disjoint for different $r$ and form a partition of $\mathcal{N}_R$). Note that for the purpose of this paper, offshore and onshore wind are considered as different resources. In addition, let $\mathcal{N}_{\texttt{FLP}}^r$ and $\mathcal{N}_{\texttt{SITE}}^r$ be the subsets of $\mathcal{N}_R^r$ selected by \texttt{FLP} and \texttt{SITE} where at least \SI{1}{MW} of capacity is deployed, respectively. Then, the screening accuracy is defined as
\begin{equation}
    \alpha_r = \frac{|\mathcal{N}^{r}_{\texttt{SITE}}\cap\mathcal{N}^{r}_{\texttt{FLP}}|}{|\mathcal{N}^{r}_{\texttt{FLP}}|}, \mbox{ } \forall r \in \mathcal{R},
\end{equation}
where $|{\mathcal{N}}|$ denotes the cardinality of set $\mathcal{N}$. First, in this table, it can be seen that the relative reduction achieved by \texttt{SITE} varies from 6\% for utility-scale PV to 62\% for distributed PV installations in the 2017 instance, with an average reduction in onshore and offshore wind sites of 38\% and 54\%, respectively. Furthermore, an overall reduction of the number of selected RES sites of up to 54\% is observed across the three considered instances. In other words, less than half of the candidate RES sites are found to be relevant in the optimal system configuration by \texttt{SITE} and subsequently passed to the \texttt{RLP}. With respect to the ability of \texttt{SITE} to identify relevant RES locations, only the distributed PV sites have a selection accuracy score below 85\%. However, the limited deployment of this technology in the solution of the proposed CEP instances enables the screening stage to properly identify over 90\% of the relevant RES sites (i.e., the ones appearing in the \texttt{FLP} solution), irrespective of the weather year considered.

\begin{table}[b]
\renewcommand{\arraystretch}{1.1}
\centering
\caption{Technology-specific sites reduction ($\gamma_r$) and screening accuracy ($\alpha_r$) of \texttt{SITE}. Number of candidate sites used in the \texttt{FLP} specified in parantheses.}
\label{tab:count}
\begin{tabular}{l|cc|cc|cc|cc}
\toprule
& \multicolumn{2}{c|}{$\mathrm{W_{on}} \hspace{1mm} (590)$} & \multicolumn{2}{c|}{$\mathrm{W_{off}} \hspace{1mm} (417)$} & \multicolumn{2}{c|}{$\mathrm{PV_{u}} \hspace{1mm} (128)$} & \multicolumn{2}{c}{$\mathrm{PV_{d}} \hspace{1mm} (605)$} \\
& $\gamma_r$ & $\alpha_r$ & $\gamma_r$ & $\alpha_r$ & $\gamma_r$ & $\alpha_r$ & $\gamma_r$ & $\alpha_r$ \\
\midrule
2016 & 0.40 & 0.94 & 0.55 & 0.85 & 0.10 & 1.00 & 0.57 & 0.54 \\
2017 & 0.37 & 0.94 & 0.55 & 0.86 & 0.06 & 1.00 & 0.62 & 0.83 \\
2018 & 0.36 & 0.93 & 0.52 & 0.85 & 0.16 & 1.00 & 0.59 & 0.59 \\
\bottomrule
\end{tabular}
\end{table}


\begin{figure}
\centering
\includegraphics[width=\linewidth]{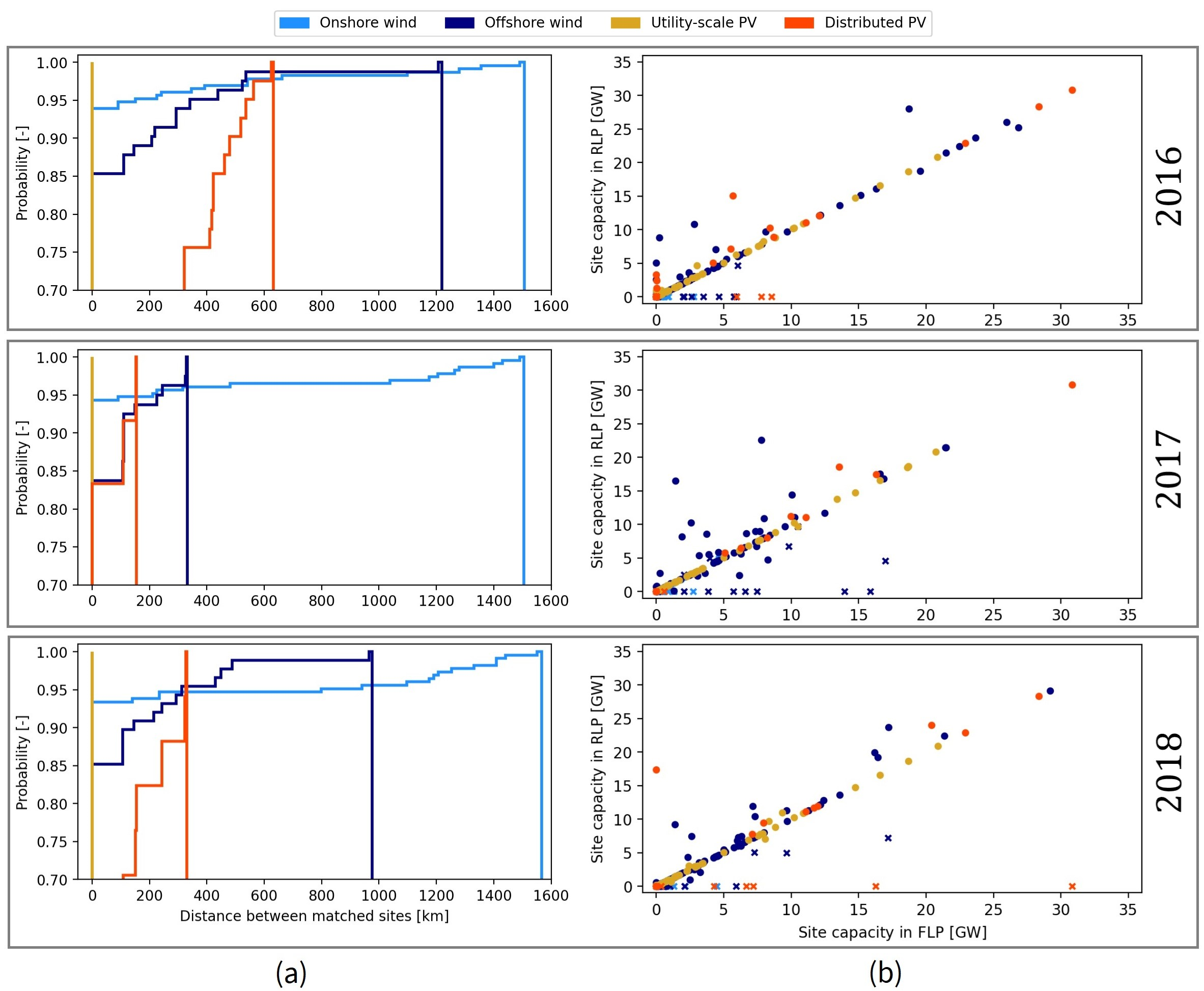}
\caption{(a) Distribution of geographical distances between pairs of sites identified via \texttt{SITE} and the \texttt{FLP}. (b) Site-specific installed capacity correlation between the \texttt{RLP} and the \texttt{FLP}.}
\label{fig:capacity_cdf}
\end{figure}

However, not all candidate RES sites found in the \texttt{FLP} solution are identified by \texttt{SITE} which selects different locations instead. For instance, when the latter is run with 2016 weather data, it fails to identify a total of 45 sites (14 onshore wind, 12 offshore wind and 19 distributed PV locations, respectively) out of 418 identified in the benchmark. Investigating how far these locations are from the ones selected by the \texttt{FLP} provides a first insight into how different the system designs associated with the two methods are. If the distances between the locations selected via \texttt{SITE} and \texttt{FLP} were found to be small, one would expect the effect of misidentifying sites to be limited, as RES patterns are usually comparable at neighboring sites. Conversely, large distances between sites identified via the two methods would often imply distinct RES patterns and could thus lead to substantial differences in the way the technologies are sized. The result of this analysis is shown in Fig. \ref{fig:capacity_cdf}a. These plots depict, for each technology and weather year, the distribution of distances (expressed in kilometres) between pairs of sites selected via the \texttt{FLP} and \texttt{SITE}, respectively. The procedure used to generate these curves is as follows. First, distances of zero are associated to the pairs of sites found by both methods ($\alpha_r$ shares in Table \ref{tab:count}). Then, each unidentified site in the \texttt{FLP} solution is matched with the geographically closest (based on the geodesic distance) location in the set of \texttt{SITE}-exclusive locations. Once two sites are paired, none of them can be subsequently matched with another. Upon pairing all unidentified sites in the \texttt{FLP} with a counterpart in \texttt{SITE}, a cumulative distribution function of technology-specific distances is plotted. It can be observed in these three plots that, without exception, the $95^{th}$ percentile of the matching distance for any of the four RES technologies falls below \SI{500}{km}. In a European context, it has been previously shown that country-aggregated wind output (usually more spatially heterogeneous than PV generation) is remarkably correlated at distances below the aforementioned threshold, especially in the North Sea basin where most onshore and offshore sites are deployed in the studied instances \cite{Malvaldi2017}. Furthermore, a maximum distance between matched sites of under \SI{1600}{km} is reported for all technologies and weather years, with the largest discrepancies being consistently observed for onshore wind locations.

\begin{table*}
\renewcommand{\arraystretch}{1.1}
\centering
\caption{Differences in system-wide capacities between \texttt{FLP} and \texttt{RLP} for various technologies and weather years. A positive value reflects more capacity installed (or higher TSCE) in the \texttt{RLP}, while a negative value indicates more capacity in the \texttt{FLP}. }
\label{tab:system_configuration}
\begin{tabular}{l|cccccccc|c}
\toprule
& $\mathrm{W_{on}}$ & $\mathrm{W_{off}}$ & $\mathrm{PV_{u}}$ & $\mathrm{PV_{d}}$ & CCGT & AC & DC & Li-Ion & TSCE \\
Year & [GW] & [GW] & [GW] & [GW] & [GW] & [TWkm] & [TWkm] & [GWh] & [b\euro] \\
\midrule
\multirow{2}{*}{2016} & -7.38 & 6.06 & 3.86 & -1.63 & 0.10 & -0.03 & -0.05 & 0.51 & 0.36 \\
& (-16.31\%) & (1.48\%) & (1.92\%) & (-1.01\%) & (0.11\%) & (-0.04\%) & (0.10\%) & (1.99\%) & (0.40\%) \\
\multirow{2}{*}{2017} & -7.94 & 4.96 & 0.60 & 7.51 & 1.01 & -2.33 & -0.16 & 1.18 & 0.44 \\
& (-12.84\%) & (1.17\%) & (0.29\%) & (7.36\%) & (2.30\%) & (-2.90\%) & (-0.29\%) & (4.07\%) & (0.52\%) \\
\multirow{2}{*}{2018} & -13.73 & 12.19 & 2.51 & -5.60 & 1.03 & 0.57 & -0.04 & -1.53 & 0.42 \\
& (-23.34\%) & (2.97\%) & (1.21\%) & (-2.99\%) & (2.54\%) & (0.65\%) & (-0.07\%) & (-3.63\%) & (0.48\%) \\
\bottomrule
\end{tabular}
\vspace{-1mm}
\end{table*}

Upon screening the candidate RES locations via \texttt{SITE}, the \texttt{RLP} is run in order to retrieve, among others, the associated installed capacities. Fig. \ref{fig:capacity_cdf}b shows, for each weather year, the correlation between installed capacities of i) the sites identified in the \texttt{FLP} and ii) the sites identified by \texttt{SITE} and sized via \texttt{RLP}. In this plot, round markers (o) denote data points associated with locations that are common to \texttt{FLP} and \texttt{SITE}, while crosses (x) represent data points corresponding to the pairs of sites matched according to the procedure described in the previous paragraph. The first remark in these plots is that in 76\% (for 2016) to 79\% (for 2018) of the cases, the installed capacities of \texttt{FLP} and \texttt{RLP} sites are matched to MW-order precision. Then, it can be observed that most of the (x) markers are situated on the bottom of the corresponding subplots. A complementary analysis of the resource signals associated with these data points suggests the existence of high-quality RES sites exploited by the \texttt{FLP}, but whose \texttt{SITE} counterparts (determined via the distance-based pairing algorithm) exhibit inferior resource quality and thus end up not being part of the \texttt{RLP} solution. In such a situation, the missing capacity, i.e., \texttt{FLP} capacity of the (x) data points in the lower part of the plot, is compensated in the \texttt{RLP} by superior power ratings at (o) sites above the trend line in Fig. \ref{fig:capacity_cdf}b.

\begin{table}[b]
\renewcommand{\arraystretch}{1.1}
\centering
\caption{Computational performance assessment of the \texttt{SM}. Numerical values represent reductions associated with the \texttt{SM} expressed in relative terms (\%) with respect to the \texttt{FLP}.}
\label{tab:performance}
\begin{tabular}{l|ccccc}
\toprule
Year & Variables & Constraints & Non-Zeros & PMR & SRT\\
\midrule
2016 & 34.54 & 34.67 & 33.48 & 41.37 & 36.56 \\
2017 & 33.31 & 33.39 & 32.22 & 40.11 & 30.90 \\
2018 & 33.72 & 33.82 & 32.67 & 39.28 & 46.57 \\
\bottomrule
\end{tabular}
\end{table}

Table \ref{tab:system_configuration} reports, for different data years and for various technologies sized within the CEP stage, the difference between the system-wide installed capacities obtained by the \texttt{FLP} and \texttt{RLP} models, respectively (positive values indicate more capacity in the latter). In the last column, it can be seen that the relative objective function difference (i.e., the TSCE) between the two CEP set-ups does not exceed 0.52\%, irrespective of the weather year considered. However, as suggested in a recent study by Neumann and Brown \cite{Neumann2021}, rather small differences in total system costs can translate into fairly distinct system configurations. In this exercise, differences of 23.3\%, 2.9\%, 1.9\% and 7.3\% are reported for onshore wind, offshore wind, utility-scale and distributed PV, respectively, between the \texttt{RLP} and the \texttt{FLP}. A closer look at the breakdown of capacities per country reveals the reasons behind such differences, as the large majority of the discrepancies observed in Table \ref{tab:system_configuration} are associated to a handful of resource-rich countries (e.g., Ireland, Italy, Spain or the UK). For instance, in 2017 and 2018, the \texttt{FLP} over-sizes onshore wind (and, thus, selects more sites) in Ireland and the UK, and uses it to supply Central Europe. Under the proposed ($\delta\tau, \xi_{\tau}^n$) set-up of the \texttt{SITE} stage, a subset of these locations are not identified (see discussion on the (x) markers in Fig. \ref{fig:capacity_cdf}a) and the associated capacity in the \texttt{FLP} is replaced in the \texttt{RLP} by a mix of offshore wind and distributed PV. Further on in Table \ref{tab:system_configuration}, transmission capacities vary within 2.9\% of the \texttt{FLP} outcome, while a maximum of 4.1\% Li-Ion storage capacity difference can be observed during the same year where distributed PV differed the most from the benchmark (i.e., 2017).

Finally, Table \ref{tab:performance} summarizes the computational performance gains (relative to the \texttt{FLP}) achieved by leveraging the \texttt{SM}. More specifically, the reductions in i) the CEP problem size (number of variables, constraints and non-zeros), ii) the peak memory requirements (PMR) and iii) the solver runtime (or SRT, taking into account the solver runtime of both the \texttt{SITE} and \texttt{RLP} stages of the \texttt{SM}) are reported. In this table, it can be observed that the proposed \texttt{SM} leads to an average CEP problem size reduction of 33\% which, in turn, enables an average PMR reduction of 40\% and runtime savings between 31\% and 46\% across the studied instances.

\section{Conclusion}\label{conclusion}
This paper proposes a method to reduce the spatial dimension of CEP frameworks while preserving an accurate representation of renewable energy sources. This is achieved via a two-stage heuristic. First, a screening stage is used to identify the most relevant sites for RES deployment among a pool of candidate locations and discard the rest. Then, the subset of RES sites identified in the first stage is used in a CEP problem to determine the optimal power system configuration. The proposed method is tested on a realistic EU case study and its performance is assessed against a CEP set-up in which the entire set of candidate RES sites is available. The method shows great promise and manages to consistently identify more than 90\% of the optimal sites while reducing peak memory consumption and solver runtime by up to 41\% and 46\%, respectively.

Capacity differences between the solutions provided by the proposed method and the benchmark observed for some weather years suggest that further work on the selection of parameters used in the first-stage siting routine would be useful. Moreover, re-casting the proposed heuristic into a more structured form, e.g., where the siting and sizing of RES assets are used as stages in a Benders-like decomposition framework, is also envisaged as a promising development avenue.

\bibliographystyle{ieeetr}
\bibliography{main}

\section*{Nomenclature}

\subsection{Abbreviations}
\begin{description}[\IEEEsetlabelwidth{$\mathrm{PV_{u}}$, $\mathrm{PV_{d}}$}]
    \item[AC, DC] AC, DC transmission links
    \item[$\mathrm{PV_{u}}$, $\mathrm{PV_{d}}$] utility-scale PV, distributed PV
    \item[TSCE] total system cost error
    \item[$\mathrm{W_{on}}$, $\mathrm{W_{off}}$] onshore wind, offshore wind
\end{description}

\subsection{Indices \& Sets}
\begin{description}[\IEEEsetlabelwidth{$\mathcal{L}_n^{+}, \mathcal{L}_n^{-}$}]
    \item[$g, \mathcal{G}$] conventional gen. tech. index and associated set
    \item[$l, \mathcal{L}$] line, set of transmission corridors, $\mathcal{L} \subseteq \mathcal{N}_B \times \mathcal{N}_B$
    \item[$\mathcal{L}_n^{+}, \mathcal{L}_n^{-}$] set of in-bound links into node $n$, with $\mathcal{L}_n^{+} = \{l \in \mathcal{L} | l = (u, n), u \in \mathcal{N}_n^{+} \}$, where $\mathcal{N}_n^{+} = \{u \in \mathcal{N}_B | (u, n) \in \mathcal{L}\}$ and set of out-bound links from node $n$, with $\mathcal{L}_n^{-} = \{l \in \mathcal{L} | l = (n, v), v \in \mathcal{N}_n^{-} \}$, where $\mathcal{N}_n^{-} = \{v \in \mathcal{N}_B | (n, v) \in \mathcal{L}\}$
    \item[$n, \mathcal{N}_B$] bus, set of buses
    \item[$m, \mathcal{N}_R$] candidate RES site and the associated set
    \item [$\mathcal{N}_R^n$, $\mathcal{N}_R^{r}$] subset of sites assigned to bus $n \in \mathcal{N}_B$, subset of sites with RES tech. $r \in \mathcal{R}$
    \item[$\mathcal{N}_{\texttt{SITE}}^n$] set of RES sites connected to bus $n \in \mathcal{N}_B$ retained in the siting stage
    \item[$\mathcal{N}_{\texttt{SITE}}^r$] set of RES sites with tech. $r \in \mathcal{R}$ retained in the siting stage
    \item[$r, \mathcal{R}$] tech. index and set of RES technologies
    \item[$s, \mathcal{S}$] storage tech. and the associated set
    \item[$t, \mathcal{T}$] time index, set of time periods
    \item[$\tau, \mathcal{T}_{\tau}$] time slice index and time slice
\end{description}

\subsection{Parameters}
\begin{description}[\IEEEsetlabelwidth{$\Delta_g^{+}, \Delta_g^{i}$}]
	\item[$\eta^{SD/D/C}_s$] efficiencies of storage tech. \textit{s}
	\item[$\phi_s$] power-to-energy ratio of storage tech. \textit{s}
	\item[$\xi_{\tau}^n$] RES supply target at node $n$, in period $\tau$
	\item[$\kappa^{0}_{l}$] initial capacity for transmission line \textit{l}
	\item[$\Bar{\kappa}_{l}$] maximum allowable installed capacity of line \textit{l}
    \item[$\kappa^{0}_{nx}$] initial capacity of tech. $x \in \{g, m, s\}$ at node \textit{n}
    \item[$\Bar{\kappa}_{nx}$] maximum allowable installed capacity of tech. $x \in \{g, m, s\}$ at node \textit{n}
    \item[$\lambda_{nt}$] electricity demand at node \textit{n} and time \textit{t}
    \item[$\pi_{nmt}$] per-unit availability of RES tech. at site $m$ (connected to bus $n$) and time \textit{t}
    \item[$\theta^{e}$] economic penalty for demand curtailment
	\item[$\theta^x_f, \theta^x_v$] fixed (FOM) and variable (VOM) operation and maintenance cost of tech. $x \in \{g, m, s, l\}$
	\item[$\omega$] weighting factor relating the length of time horizon \begin{math}\mathcal{T}\end{math} to the annualized costs of technologies
	\item[$\zeta^x$] annualized investment cost of tech. $x \in \{g, m, s, l\}$
\end{description}

\subsection{Variables}
\begin{description}[\IEEEsetlabelwidth{$p_{nxt} \in \mathbb{R_+}$}]
\item[$K_{l} \in \mathbb{R_+}$] installed capacity of transmission line \textit{l}
\item[$K_{nx} \in \mathbb{R_+}$] installed cap. of tech. $x \in \{g, m, s\}$ at bus \textit{n}
\item[$p^{C/D}_{nst} \in \mathbb{R_+}$] flows of storage tech. \textit{s} at bus \textit{n} and time \textit{t}
\item[$p_{lt} \in \mathbb{R}$] power flow over line \textit{l} at time \textit{t}
\item[$p_{nxt} \in \mathbb{R_+}$] feed-in of gen. tech./site $x \in \{g, m\}$ at bus \textit{n} and time \textit{t}
\item[$p^{e}_{nt} \in \mathbb{R_+}$] unserved demand at bus \textit{n} and time \textit{t}
\end{description}

\end{document}